# Spin–orbit interaction in Pt or Bi$_2$Te$_3$ nanoparticle-decorated graphene realized by a nanoneedle method

T. Namba[1], K. Tamura[1], K. Hatsuda[1], T. Nakamura[2], C. Ohata[1], S. Katsumoto[2], J. Haruyama[1,2*]

[1]*Faculty of Science and Engineering, Aoyama Gakuin University, 5-10-1 Fuchinobe, Sagamihara, Kanagawa 252-5258, Japan*
[2]*Institute for Solid State Physics, The University of Tokyo, 5-1-5 Kashiwanoha, Kashiwa, Chiba 277-8581, Japan*
To whom correspondence should be addressed. E-mail: J-haru@ee.aoyama.ac.jp

**The introduction of spin–orbit interactions (SOIs) and the subsequent appearance of a two-dimensional (2D) topological phase are crucial for voltage-controlled and zero-emission energy spintronic devices. In contrast, graphene basically lacks SOIs due to the small mass of the carbon atom, and appropriate experimental reports for SOIs are rare. Here, we control small-amount (cover ratios < 8%) random decoration of heavy nanoparticles [platinum (Pt) or bismuth telluride (Bi$_2$Te$_3$)] onto mono-layer graphene by developing an original nanoneedle method. X-ray photoelectron spectra support low-damage and low-contamination decoration of the nanoparticles, suggesting the presence of Bi–C and Te–C coupling orbitals. In the samples, we find particle-density-dependent non-local resistance ($R_{NL}$) peaks, which are attributed to the (inverse) spin Hall effect (SHE) arising from SOI with energies as large as ~30 meV. This is a larger value than in previous reports and supported by scanning tunneling spectroscopy. The present observation should lead to topological phases of graphene, which can be introduced by random decoration with controlled small amounts of heavy nanoparticles, and their applications.**

For the realization of functional spintronic devices, significant attention has been focused on spin–orbit interactions (SOIs),[1] which allow control over spins by applied external electric fields. On the other hand, graphene, a 2D atom-thin carbon layer, basically lacks SOIs. When graphene is sufficiently isolated from the influence of the substrate, long spin relaxation times ($\tau_s$) and large mean free paths of electrons are provided. If SOIs could be introduced without suppressing these properties, graphene with SOIs could yield various quantum phenomena [e.g., 2D or one-dimensional topologically insulating (TI) states[4-10,18,23]] and their practical applications for voltage-controlled spintronic devices. Many papers in the literature have predicted them theoretically.

Recently, the experimental challenge to introduce SOIs into graphene has been met by a number of methods [e.g., surface decoration by (1) right-mass adatoms[2,3] or (2) heavy nanoparticles,[11-13] and (3) using heavy wires[14-16] or substrates]. Nevertheless, they suffer from a lack of appropriate experimental results. One of the reasons for (1) and (2) is that precise control over the small amount of decoration (e.g., coverage ratio < 10%) without causing damage or contamination is difficult. Moreover, a large amount decoration (e.g., > 10%) leads to various parasitic phenomena (e.g., spin absorption,[16,21] phase interference of electron spin waves or dephasing,[2] or intervalley scattering) and these can obstruct the observation of pure SOI. Therefore, the small-amount and damageless decoration of nanoparticles onto graphene is crucial.

On the other hand, heavy adatom (nanoparticle) decoration theoretically leads to SOIs in graphene, preserving the *sp$^2$* bond character of graphene and mediating diverse SOIs through electron tunneling onto and off the adatom *p,d,f*-orbitals, depending on the position of the adatoms on the hexagonal carbon lattice of graphene. There is much theoretical literature for this. For instance, SOI energy ($E_{SO}$) ~40 meV for a concentration of ~0.1 lead (Pb) adatom/carbon of graphene was theoretically predicted with a dependence on adatom density and position.[17] An $E_{SO}$ > 200 meV with a robust 2D TI state was even predicted for $5d_{xz,\,yz}$-adatom [e.g., osmium (Os)] decorated graphene with coverage as small as 1% due to hybridization of the partially filled $d$-orbital SOI impurity bands with the Dirac state of graphene.[18] Random and small-amount decoration (coverage << 10%) can even stabilize topological phases.[23] Moreover, the decoration of graphene by a small amount of nanoparticles can maintain ballistic electron transport, resulting in room-temperature detection of spin current. Although these features are advantageous, the predicted large $E_{SO}$ has not been obtained experimentally. In this study, we reveal how Pt or $Bi_2Te_3$ nanoparticles, which have large SOIs but also 3D TI states for the latter, introduce large SOIs into graphene by small-amount and damageless random decoration developed using an innovative nanoneedle method.

In the present experiments, Pt or $Bi_2Te_3$ nanoparticles with diameters of 3 ~ 50 nm (Sigma Aldrich Inc.) are decorated onto chemical vapor deposition (CVD)-grown mono-layer graphene surfaces formed into a multiple Hall bar shape (Figs. 1a and 1b; area for [width($w$) = 2 μm] × [length($L$) = 5 μm]). The high quality of the mono-layer graphene has been confirmed by Raman spectra and X-ray photoelectron spectroscopy (XPS).

For this experiment, we develop a specific tool, the so-called nanoneedle method (inset of Fig. 1c; Saito Medical Instruments Inc.), to complete the clean and damageless decoration of a graphene surface with nanoparticles. Before decoration, an acetone solution containing the nanoparticles is ultrasonicated for 24 hours in order to reform the nanoparticles to a smaller diameter. Dropping an acetone droplet from the narrow top of the needle, which has an inner pore diameter of ~50 μm, makes it possible to control the small particle density ($D$) on the narrow graphene surface formed to the multiple Hall bar pattern. Examples of controlled random nanoparticle-decoration with three different values of $D$ are shown in Fig. 1 (c – e) for $D$ ~ 4/100$^2$ nm$^2$ (coverage ~ 3%), $D$ ~ 10/100$^2$ nm$^2$ (coverage ~8%), and $D$ ~ 23/100$^2$ nm$^2$ (coverage ~20%), respectively. The typical $D$ used for the present nanoparticle decoration of the graphene Hall bar corresponds to (d). After the decoration, the samples are annealed at 400 °C for 10 minutes under a high vacuum to activate the chemical combination of the nanoparticles with the carbon atoms of graphene, inducing damageless and contaminationless decoration.

Typical XPS spectra of this sample also demonstrated small peaks arising from Bi–C coupling (~282 eV), $Bi4f_{5/2}$ and $Te3d_{5/2}$ orbitals in isolated Bi and Te (i.e., without oxidization) (~163 eV and 572 eV), and Te–C coupling (~574 eV). These suggest that a tunneling current can be caused through the $d$, $f$-orbitals of the nanoparticles coupled with the graphene Dirac state, as mentioned above.[17,18] Such small peaks for Bi–C and Te–C coupling could not be observed in samples fabricated by other decoration methods (e.g., sputtering and evaporation), which damage and contaminate the graphene surface. It suggests that the nanoneedle method in our experiments can effectively decorate graphene with adatoms with little disturbance.

Figure 2 shows the room-temperature measurement results of the non-local resistance ($R_{NL}$) of graphene decorated with Pt-nanoparticles ($D$ ~ 10/100$^2$ nm$^2$ for coverage ~8%), measured for the pattern in Fig 1 with the back gate electrode consisting of Au/Ti. Ohmic resistances have been subtracted. Figure 2a plots $R_{NL}$ as a function of the back gate voltage ($V_{bg}$) for electrode pair 2–7 ($R_{27}$), which is located at the closest position to the constant current flow $I_{18}$. Compared with the $R_{NL}$ peak in bare graphene without Pt particles ($R_{bare}$), the $R_{NL}$ peak becomes much larger after particle decoration. The peak amplitude is larger than those in previous reports for nanoparticle-decorated graphene. Figure 2b exhibits the dependence of the $R_{NL}$ vs. $V_{bg}$ relationship on the distance ($L$) between electrode pair 1–2 and the electrode pairs used to observe $R_{NL}$. As $L$ increases, the height of the $R_{NL}$ peak also quickly decreases. The relationship is clearly evident in Fig. 2c The $R_{NL}$ peak height exponentially decays with increasing $L$. It is well fitted by the following diffusion equation (the dotted line in Fig. 2b),[11]

$$R_{NL} = \frac{1}{2}\gamma^2\rho\frac{w}{\lambda_s}e^{-L/\lambda_s}, \qquad (1)$$

where $\lambda_S$, defined as $\lambda_S (= \lambda_{SO}^2/w)$, is the spin relaxation length ($\lambda_{SO}$ is the SO relaxation length), and $\gamma$ and $\rho$ are the spin Hall angle and resistivity of the sample, respectively. Using a fixed $w = 2$ μm, the best fit provides $\gamma \sim 0.4$ and $\lambda_S \sim 0.8$ μm. These values suggest the presence of a much stronger SOI [spin Hall effect (SHE)] compared with previous reports for nanoparticle-decorated graphene. The $\gamma$ value is larger than that in Pt thin films (e.g., ~0.37). The Ohmic contribution is negligible when $L > \lambda_S \sim 0.8$ μm (i.e., in the present case). Indeed, it can be estimated to be as small as $e^{-\pi\lambda_s/w} \approx e^{-58}$ using $\lambda_S \sim 0.8$ μm and $w = 2$ μm.[24] The $L$ and $w$ dependence of $R_{NL}$ in much smaller regions (i.e., $L, w \ll 1$ μm) could not be measured, because precise control of the same small $D$ values could not be obtained in these regions, even using nanoneedles.

To reconfirm SOIs, Larmor spin precession (the Hanle effect) is measured by applying an in-plane magnetic field ($B_{//}$) at low temperature (T = 1.5K) (Fig. 2d). $R_{NL}$ exhibits two different oscillatory behaviors as $B_{//}$ changes. The oscillatory behavior observed around $B_{//} = 0$ is much sharper than previous reports for nanoparticle-decorated graphene, while moderate oscillatory behavior with an asymmetric property for $\pm B_{//}$ regions is observed at high $B_{//}$ (i.e., $0 \ll B_{//} < \sim\pm3$). In a region of much higher $B_{//}$, the oscillatory behavior disappears. The sharp oscillatory behavior is strong evidence for the presence of SOIs, because only SHE can lead to such behavior. The $R_{NL}$ vs. $B_{//}$ curve can be fitted by the following equation through all $B_{//}$ region, [11]

$$R_{NL} = \frac{1}{2}\gamma^2\rho w Re\left[(\sqrt{1+i\omega_B\tau_s}/\lambda_s)e^{-(\sqrt{1+i\omega_B\tau_s}/\lambda_s)L}\right], \qquad (2)$$

where $\omega_B = \Gamma B_{//}$ is the Larmor frequency with $\Gamma$, which is the gyromagnetic ratio for electrons. The best fit to the curve for $0 \ll B_{//} < \sim +3$ gives $\gamma \sim 0.4$, $\lambda_S \sim 0.9$ μm, and $\tau_s \sim 18$ ps and for $0 \gg B_{//} > \sim -3$ gives $\gamma \sim 0.38$, $\lambda_S \sim 1.2$ μm, and $\tau_s \sim 16$ ps (blue dashed line), when different fitting parameters are employed for $\pm B_{//}$ regions which show the strong asymmetry. These values are almost consistent with the values obtained from Fig. 2d and, hence, support the presence of strong SOIs, particularly around $B_{//} = 0$. Although $B_{//}$ dependence was measured only at T = 1.5K in the present work, it is expected that the strong SHE observed at T = 300K should lead to appearance of this Hanle effect (oscillatory $R_{NL}$ behavior) even up to T = 300K.

When the Elliott–Yafet (EF) mechanism, which is a dominant factor for spin relaxation in conventional graphene, is assumed, $E_{SO}$ is given by $\tau_s = (\varepsilon_F/E_{SO})^2\tau_p$, where $\varepsilon_F$ is the Fermi energy and $\tau_p$ is the momentum relaxation time. $E_{SO}$ is estimated to be ~20 meV, using $\tau_s \sim 16$ ps mentioned above and $\varepsilon_F$ at $n = \sim 3 \times 10^{12}$ cm$^2$ of our samples. On the other hand, the D'yakonov–Perel' mechanism has been experimentally reported only in some graphene (e.g., with a heavy substrate like tungsten). Because the present graphene is decorated with very small $D$ and without damage or contamination, the EF mechanism should be appropriate. Indeed, the EF mechanism has been previously reported in small-$D$-hydrogenated graphene[2] and heavy-nanoparticle-decorated graphene.[11]

Figure 3 demonstrates the room-temperature observation of the $R_{NL}$ peaks for graphene decorated with $Bi_2Te_3$ nanoparticles (*D* for coverage ~8%). Figures 3a and 3b confirm the large $R_{NL}$ peaks, which quickly reduce with increasing *L*. The best fit to Fig. 3b using Eq. (1) gives $\gamma \sim 0.45$ and $\lambda_S \sim 0.75$ μm, which suggest the presence of SOI stronger than that in Pt-particle-decorated graphene.

Figure 3c shows the *B*-dependence of the $R_{27}$ peak of Fig. 3a. The oscillatory behavior observed around $B_{//} = 0$ is much sharper than that in Fig. 2d, while moderate oscillatory behavior is observed in the high $B_{//}$ region. The best fit through all $B_{//}$ regions using eq. (2) suggests stronger SOI and Larmor spin precession with $\gamma \sim 0.48$, $\lambda_S \sim 0.9$ μm, and $\tau_s \sim 18$ ps for $0 << B_{//} < \sim +3$ and $\gamma \sim 0.47$, $\lambda_S \sim 0.7$ μm, and $\tau_s \sim 13$ ps for $0 >> B_{//} > \sim -3$, when different fitting parameters are used for $\pm B_{//}$ regions (blue dashed line). $E_{SO}$ is also estimated to be ~30 meV from these parameters, while the data fit to only the sharp curve around $B_{//} = 0$ gives much larger $E_{SO}$ (e.g., maximum ~50 meV). In contrast, $R_{NL}$ does not exhibit any oscillatory behavior under out-of-plane $B$ ($B_\perp$). These findings also support the presence of SOIs in $Bi_2Te_3$-nanoparticle-decorated graphene. As the temperature increases, this sharp $R_{NL}$ peak is reduced, while it still remains even at T = 20K (Fig. 3d).

Scanning tunneling spectroscopy (STS) spectra reconfirm the observed $E_{SO}$ value. Figure 4a shows an STS spectrum of graphene decorated with $Bi_2Te_3$ nanoparticles (*D* for coverage ~8%). An evident gap of ~20 meV is confirmed close to the nanoparticle, while no evident gap is confirmed away from the nanoparticle. This is almost consistent with the observation and analysis mentioned above.

Figure 4b demonstrates $R_{NL}$ peak values as a function of *D* for Pt and $Bi_2Te_3$. $R_{NL}$ peak values for both cases monotonically increase in regions below $D \sim 10/100^2$ nm$^2$ as *D* increases. On the other hand, above $D \sim 10/100^2$ nm$^2$, the increase ratio is reduced (i.e., saturate or provide a lower slope value).

We discuss the origins for the observed large SOI [(inverse) SHE, $\gamma$]. XPS observation confirmed that the *3d,4f*-orbitals of the $Bi_2Te_3$ nanoparticles have hybridized with the graphene Dirac state. Some models based on such *p, d,* and *f* orbital hybridizations have been proposed in adatom-decorated graphene, as mentioned above.[17-20,22.23] The electron tunneling between two carbon atoms via the *p, d,* and *f* orbitals of the adatoms opens up additional channels for hopping in graphene. The SO coupling between the orbitals of the adatom induces either intrinsic SOI (by conserving the spin) or Rashba-like SOI (by flipping the spin) through the tunneling channels, depending on the size, chemical condition, and position of the adatoms.

Here, large $\gamma$ values (e.g. ~0.2 – 0.5) are reported and the mechanisms are discussed in lightly hydrogenated graphene,[2] heavy-nanoparticle-decorated graphene,[11-13] and graphene/Pt wire,[14-16] as follows; e.g., (1) the strong Fermi-energy dependence of the density of states and its sharp sensitivity to adsorbed nanoparticles (i.e., SOI enhances the SHE via skew scattering of charge carriers in the resonant regime), (2) the efficient spin injection into graphene in combination with shunting-current suppression, and (3) hybridization of the partially filled *d*-orbital impurity bands of the adatoms with the Dirac state of graphene. In contrast, the present $\gamma$ values (SOI) are even larger than these values. One of differences from previous experiments is the low-damage and low-contamination decoration of graphene with $Bi_2Te_3$ nanoparticles realized by our nanoneedle method. This might be associated with the origins for the observed large $\gamma$, because the robust chemical bonds derived from them induce strong tunneling current between the nanoparticles and graphene. Indeed, confirmation of the clear chemical bonds has merely been reported in XPS in other previous reports. Moreover, they have not been confirmed in graphene decorated with nanoparticles by other methods (e.g., sputtering and evaporation), which risk the possibility of causing more damage and contamination to the graphene surface, than in our experiments. On the other hand, the best data fit as mentioned above (e.g. in *L* dependence) may be overestimated, because *w* dependence could not be measured due to the difficulty over the precise control of the same small *D* for $w << 1$ μm.

In contrast to the advantages for the introduction of a large SOI, at least the following three demerits can be considered due to the nanoparticle decoration, particularly with large $D$ values; (1) spin absorption effect,[16,21] (2) spin-phase interference or dephasing,[3] and (3) intervalley scattering. For (1), for example, CuBi wires placed on the Cu channel between the two Py contacts absorbed pure spins injected from the Py electrode, depending on Bi concentration. In (2), metallic particles can cause interference of phases of the electron spin waves or their dephasing, leading to fluctuations in the device resistance. However, these disadvantages are not dominant for the present small coverage ratio ($<$ ~8 %) within a random position. Indeed, $R_{NL}$ peak values show linear relationships at $D \leq$ ~$10/100^2$ nm$^2$ ($<$ ~8%) in Fig. 4b, while the increase ratio is reduced at $D >$ ~$10/100^2$ nm$^2$, suggesting appearance of the parasitic effects as mentioned above. Small $D$ leads to stable SOI by avoiding such parasitic effects.[18,23]

The logarithmic temperature dependence of non-local conductance ($G_{NL}$) dip values of Bi$_2$Te$_3$-nanoparticle-decorated graphene is shown for two typical samples within these two different $D$ regimes in Fig. 4c. In the low-$D$ regime ($<$ ~8%), $G_{NL}$ is almost independent of the temperature, while it shows a linear decrease with decreasing temperature and saturation at low temperature ($<$ 10K) in the high-$D$ regime ($>>$ ~8%). The latter can be understood by weak localization (WL) arising from the constructive electron–wave phase interference in a diffusive charge transport regime.[2] In WL, constructive phase interface of two electron waves (i.e., interference by the same phases) occurs at the electron injection point of the sample and, thus, the electron localizes around there, resulting in a resistance maximum. In Fig. 4c, the linear region existing at high temperatures means dephasing of this phase interference due to electron–electron interaction with increasing temperature. In contrast, the saturation regime at low temperature suggests the appearance of conductance which is independent of a temperature change, such as spin scattering by magnetic impurities. When magnetic impurities are absent, this saturation region can be attributed to SOI scattering. Thus, the high critical temperature (~10K) between the linear and saturation regions implies the presence of either strong spin or SOI scattering. This suggests that the $R_{NL}$ peaks observed in the high-$D$ region are not due to pure SOI, while the weak temperature dependence, which is consistent with the absence of the negative magnetoresistance for $B_\perp$ in Fig. 3c, suggests that those in the low-$D$ region can be attributed to pure SOI. These demonstrate an example where the abovementioned disadvantages become dominant only in high-$D$ regions or under great damage to graphene.

In conclusion, we controlled small-amount (coverage ratios $<$ 8%) random decoration of heavy nanoparticles (Pt or Bi$_2$Te$_3$) onto mono-layer graphene by developing an original nanoneedle method. XPS spectra suggested little damage and low contamination on the graphene surface, leading to the emergence of the coupling orbitals for Bi–C and Te–C. In the samples, we observed $D$-dependent $R_{NL}$ peaks at room temperature, which suggested the presence of SOI energies (~30 meV) which were estimated from the data fits, and (inverse) SHE. This value, supported by STS, was larger than those in previous reports. The present nanoneedle method and the observed large SOI are highly expected to yield topological phases of graphene, when a much smaller sample size is employed and decoration methods for much smaller $D$ are improved further to yield a robust helical edge spin mode.

The authors thank Y. Shimazaki, T. Yamamoto, S. Tarucha, T. Enoki, M. Dresselhaus, J. Shi, and P. Kim for their technical contributions, fruitful discussions, and encouragement. The work at the Aoyama Gakuin University was partly supported by a grant for private universities and a Grant-in-Aid for Scientific Research (15K13277) awarded by MEXT.

**References**

(1) J. Ryu, M. Kohda, and J. Nitta, *Phys. Rev. Lett.* **116**, 256802 (2016).
(2) J. Balakrishnan, G. Koon, M. Jaiswal, A. H. Castro Neto, B. Özyilmaz, *Nat. Phys.* **9**, 284 (2013).
(3) T. Kato, J. Kamijo, T. Nakamura, C. Ohata, S. Katsumoto and J. Haruyama, *Royal Society of Chemistry Advances* **6**, 67586 (2016).
(4) A. Roth, C. Brüne, H. Buhmann, LW. Molenkamp, J. Maciejko, XL Qi, SC. Zhang, *Science* **325**, 294 (2009).
(5) C. Brune, A. Roth, H. Buhmann, E. M Hankiewicz, L. W Molenkamp, J. Maciejko, X.-L. Qi, S.-C. Zhang, *Nature Phys*. **8**, 485 (2012).
(6) C.L. Kane and Mele, E. J. *Phys. Rev. Lett.* **95**, 146802 (2005).
(7) C.L. Kane and E. J. Mele, E. J. *Phys. Rev. Lett.* **95**, 226801 (2005).
(8) V. Mourik, K. Zuo, S. M. Frolov, S. R. Plissard, E. P. A. M. Bakkers, L. P. Kouwenhoven, *Science* **336**, 1003 (2012).
(9) S. M. Albrecht, A. P. Higginbotham, M. Madsen, F. Kuemmeth, T. S. Jespersen, J. Nygård, P. Krogstrup & C. M. Marcus, *Nature* **531**, 206 (2016).
(10) M.T. Deng, S. Vaitiekėnas, E. B. Hansen, J. Danon, M. Leijnse, K. Flensberg, J. Nygård, P. Krogstrup & C. M. Marcus, *Science* **354**, 1557 (2016).
(11) J. Balakrishnan G. K.W. Koon, A. Avsar, Y. Ho, J. H. Lee, M. Jaiswal, S.-J. Baeck, J.-H. Ahn, A. Ferreira, M. A. Cazalilla *et al*., *Nature Commun.* **5**, 4748 (2014).
(12) D. Van Tuan J. M. Marmolejo-Tejada, X. Waintal, B. K. Nikolić, S. O. Valenzuela, and S. Roche. *Phys. Rev. Lett.* **117**, 176602 (2016).
(13) Y. Wang X. Cai, J. R.-Robey, and M. S. Fuhrer, *Phys. Rev. B* **92**, 161411(R) (2015).
(14) W. S. Torres, J. F. Sierra, L. A. Benítez, F. Bonell, M. V. Costache and S. O. Valenzuela, *2D Mater.* **4**, 041008 (2017).
(15) T. Kimura, Y. Otani, T. Sato, S. Takahashi, and S. Maekawa, *Phys. Rev. Lett.* **98** 156601 (2007).
(16) L. Vila, T. Kimura, and Y. Otani, *Phys. Rev. Lett.* **99** 226604 (2007).
(17) L. Brey, *Phys. Rev. B* **92**, 235444 (2015).
(18) J. Hu, J. Alicea, R. Wu, and M. Franz, *Phys. Rev. Lett.* **109**, 266801 (2012).
(19) W.K. Tse, Z. Qiao, Y. Yao, A. H. MacDonald, and Q. Niu, *Phy. Rev. B* **83**, 155447 (2011).
(20) Z. Qiao, S. A. Yang, W. Feng, W.-K. Tse, J. Ding, Y. Yao, J. Wang, and Q. Niu, *Phys. Rev. B* **82**, 161414(R) (2010).
(21) Y. Niimi, Y. Kawanishi, D. H. Wei, C. Deranlot, H. X. Yang, M. Chshiev, T. Valet, A. Fert, and Y. Otani, *Phys. Rev. Lett.* **109**, 156602 (2012).
(22) H. Zhang, C. Lazo, S. Blügel, S. Heinze, and Y. Mokrousov, *Phys. Rev. Lett.* **108**, 056802 (2012).
(23) H. Jiang, Z. Qiao, H. Liu, J. Shi, Q. Niu, *Phys. Rev. Lett.* **109**, 116803 (2012).
(24) D. A. Abanin, A. V. Shytov, L. S. Levitov, and B. I. Halperin, *Phys. Rev. B* **79**, 035304 (2009).

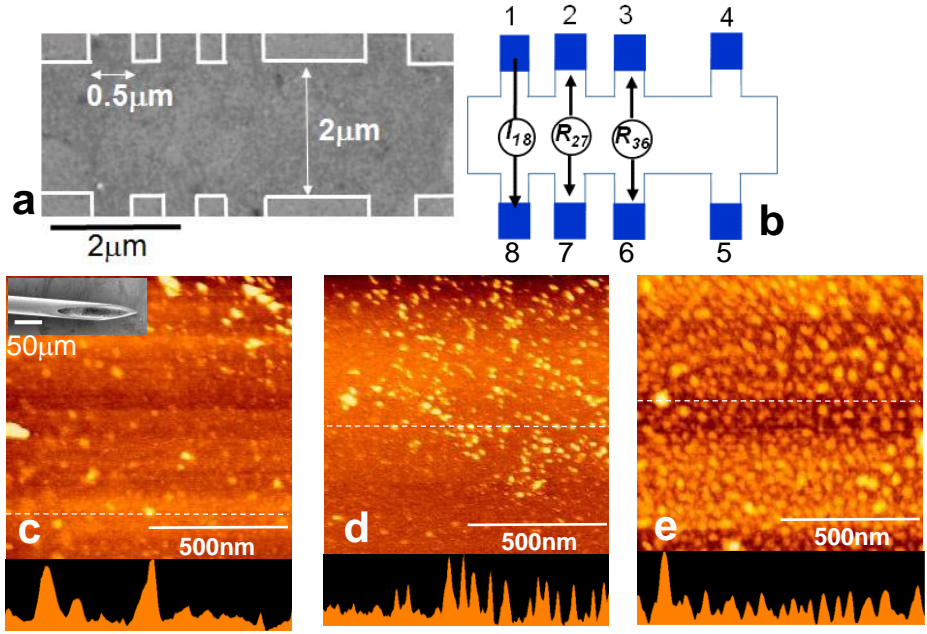

**FIG. 1 (a)(b)** SEM and schematic top views of graphene formed into a multiple Hall bar pattern used for non-local resistance ($R_{NL}$) measurements. **(c – e)** AFM images of the $Bi_2Te_3$ nanoparticle decorated on graphene using a nanoneedle method (**inset of (c)**) with three different densities (*D*) ; (c) – (e) for $D \sim 4/100^2$ nm$^2$ (coverage ~3%), $D \sim 10/100^2$ nm$^2$ (coverage ~8%), and $D \sim 23/100^2$ nm$^2$ (coverage ~20%), respectively. White dotted lines of top views (upper panels) correspond to the positions measured for cross-sectional views (lower panels).

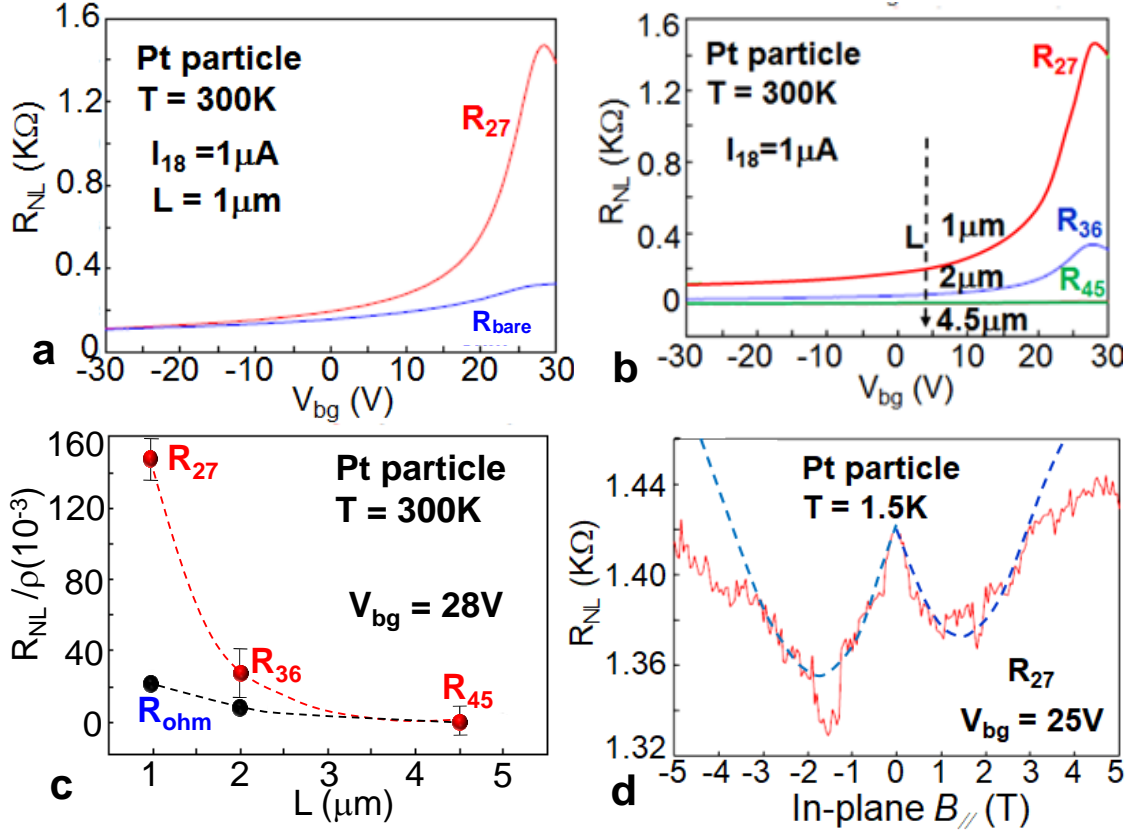

**FIG. 2 (a)** Non-local resistance ($R_{NL}$) vs. back gate voltage ($V_{bg}$) measured in Pt-nanoparticle-decorated graphene ($D \sim 10/100^2$ nm$^2$), using a constant current flow at electrode pair 1–8 ($I_{18}$) and a $R_{NL}$ electrode pair 2–7 ($R_{27}$) in the Fig. 1(b)-pattern. Ohmic resistances have been subtracted. **(b, c)** Distance ($L$) dependence of $R_{NL}$, using $R_{27, 36, 45}$ in the Fig. 1(b)-pattern. $L$ is the distance between $I_{18}$ electrode pair and individual $R_{NL}$ pair. **(d)** In-plane $B$ ($B_{//}$) dependence of the $R_{NL}$ peak of (a). Dashed lines in (c) and (d) are the best fits by eqs. (1) and (2), respectively. For (d), different fitting parameters are used for $+B_{//}$ and $-B_{//}$ regions.

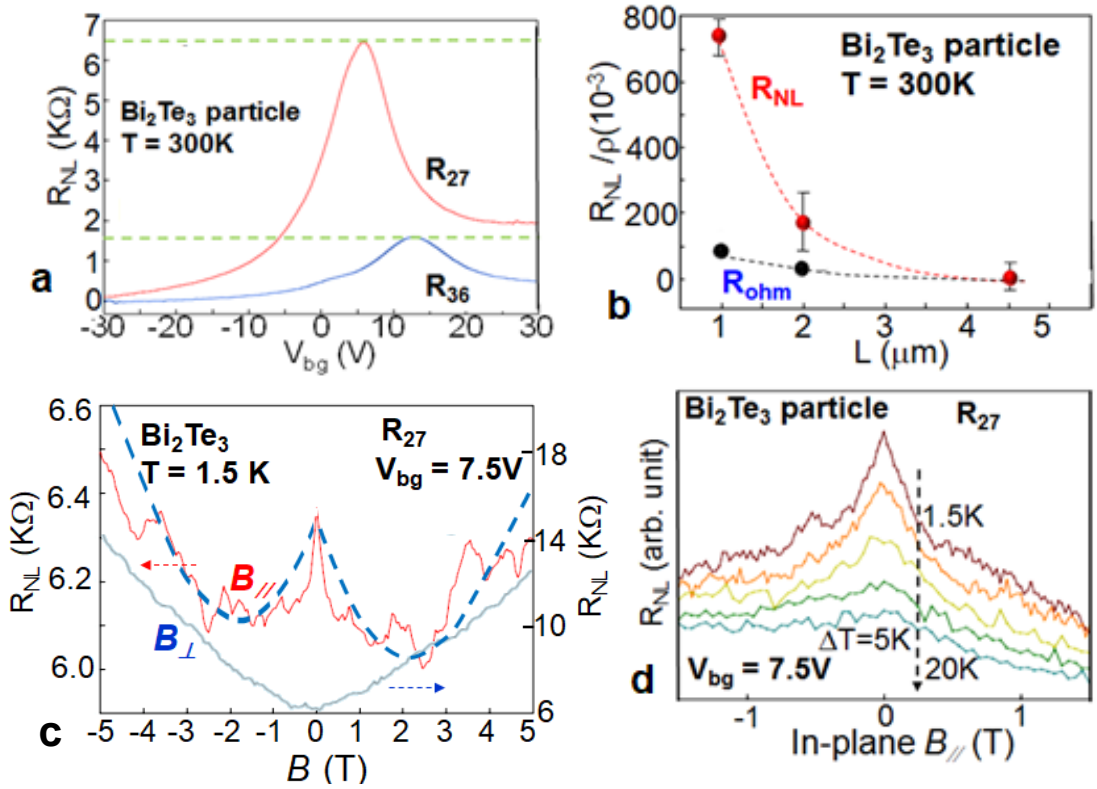

**FIG. 3** (a) Non-local resistance ($R_{27}$ and $R_{36}$) vs. $V_{bg}$ measured in a $Bi_2Te_3$-nanoparticle-decorated graphene ($D \sim 10/100^2$ nm$^2$), using the Fig. 1(b)-pattern. Ohmic resistances have been subtracted. Green dotted lines correspond to $R_{NL} = h/4^n e^2$ for $n$ = 1 and 2. **(b)** Distance dependence of $R_{NL}$ and $R_{ohm}$, using $R_{27, 36, 45}$ in the Fig. 1(b)-pattern. **(c)** In- and out-of-plane $B$ ($B_{//}$ and $B_\perp$) dependence of the $R_{27}$ peak of (a). Dashed lines in (b) and (c) are the best fits by eqs. (1) and (2), respectively. For (c), different fitting parameters are used for $+B_{//}$ and $-B_{//}$ regions. **(d)** Temperature dependence of the $B_{//}$-derived $R_{27}$ peak in (c).

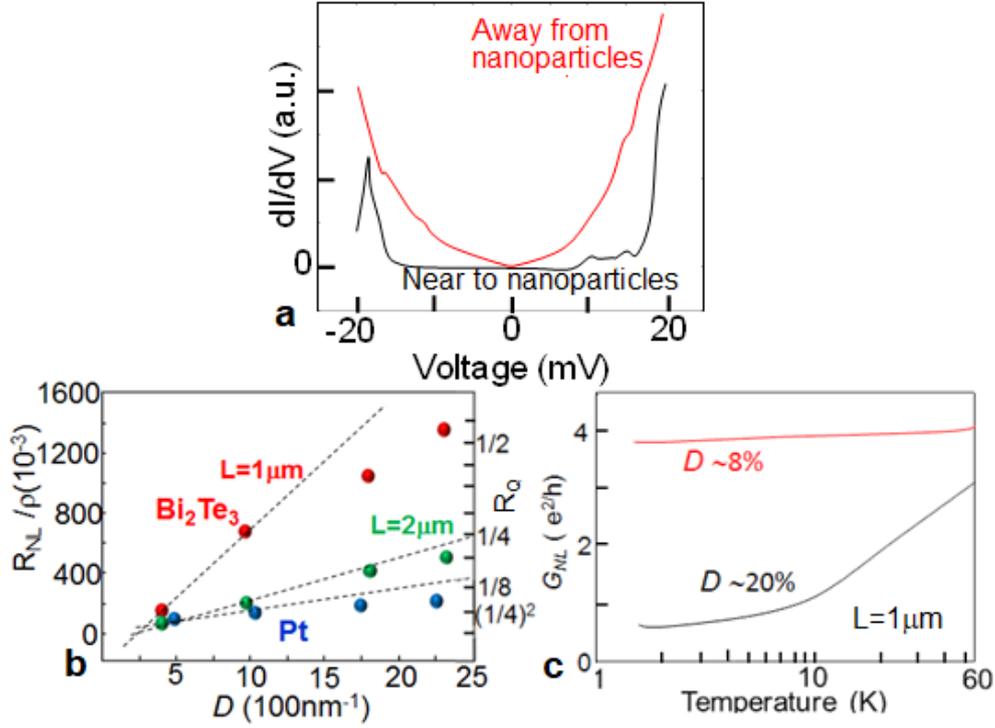

**FIG. 4 (a)** Typical STS spectra of $Bi_2Te_3$-nanoparticle-decorated graphene ($D \sim 10/100^2$ nm$^2$) at T = 400 mK and $V_{bg}$ = 8 V, taken around a $Bi_2Te_3$ nanoparticle (within a few 10 nm) and ~3 μm away from the nanoparticle. **(b)** $R_{NL}$ peak values as a function of $D$ for Pt (blue plot for $L$ = 1 μm) and $Bi_2Te_3$ (red and green plots for $L$ = 1 μm and 2 μm, respectively). Dotted lines are just a guide for the eyes. **(c)** Logarithmic temperature dependence of non-local conductance ($G_{NL}$) dip values for typical two $Bi_2Te_3$-nanoparticle-decorated graphenes within two different $D$ regimes.